\documentclass{ws-procs9x6}

\usepackage{colordvi}

\begin{document}

\title{VERTEX COUPLINGS IN QUANTUM GRAPHS: APPROXIMATIONS BY SCALED SCHR\"ODINGER OPERATORS}

\author{PAVEL EXNER}

\address{Doppler Institute for Mathematical Physics and Applied
Mathematics, \\ Czech Technical University,
B\v rehov{\'a} 7, 11519 Prague, Czech Republic, and}
\address{Department of Theoretical Physics, Nuclear Physics
Institute, \\ Czech Academy of Sciences,
25068 \v{R}e\v{z} near Prague, Czech Republic\\
E-mail: exner@ujf.cas.cz}

\begin{abstract}
We review recent progress in understanding the physical meaning of quantum graph models through analysis of their vertex coupling approximations.
\end{abstract}


\bodymatter

\section{Introduction}\label{s1:intro}

Quantum graphs attracted a lot of interest recently. There are several reasons for that. On one hand these models are useful as descriptions of various structures prepared from semiconductor wires, carbon nanotubes, and other substances. On the other hand they provide a tool to study properties of quantum dynamics in situations when the system has a nontrivial geometrical or topological structure.

Quantum graph models contain typically free parameters related to coupling of the wave functions at the graph vertices, and to get full grasp of the theory one has to understand their physical meaning. A natural approach to this question is to investigate ``fat graphs'', that is, systems of thin tubes built over the skeleton of a given graph, and to analyze its limit as the tube thickness tends to zero.

While simple at a glance, the problem is in fact rather difficult and its understanding is being reached through a long series of works. The aim of the present paper is to review some recent achievements in this area. We present this survey in a non-technical way referring for detailed proofs and a wider background to the literature, in particular, to our recent papers [\refcite{cet10a, ep09}]. Having said that it is important to stress that we will formulate the problem and the results in a fully rigorous way.

\section{Quantum graphs}

\subsection{A bit of history}

The quantum graph concept was born in early days of quantum mechanics being first suggested in the 1930's by Linus Pauling as a model of aromatic hydrocarbons, and worked out later by Ruedenberg and Scherr [\refcite{rs53}]. Then, as it sometimes happen in the history of science, it was happily forgotten.

In a sense it might be surprising because the idea of a quantum particle living on a graph is theoretically attractive, however, it was not enough and for three decades quantum graph models enjoyed the status of an obscure textbook example. This changed in the eighties, when the diminishing size of structures produced in solid-state-physics laboratories reached the state when the electron transport in them became dominantly ballistic and quantum graphs suddenly reemerged as a useful model.

The list of physical system to which these methods can be applied kept expanding. At the beginning it included microstructures fabricated from semiconductor or metallic materials, later carbon nanotubes were added. It is worth mentioning, however, that the same technique can be used also to investigation of electromagnetic phenomena in large network-type structures [\refcite{hul04}], at least as long stationary situation is considered.

Quantum dynamics of a particle confined to a graph can mean various things, of course. Typically one considers a nonrelativistic  situation described by a \emph{Schr\"odinger operator} supported by the graph. Often the motion is free but in other situations one adds potentials corresponding to external electric or magnetic fields, spin degrees of freedom, etc. Graphs can support also \emph{Dirac operators}. Such a model, too, was for a long time regarded as a theoretician toy and attracted a limited attention only [\refcite{bh03,bt90}]. The situation changed dramatically two or three years ago with the discovery of graphene in which electron behave effectively as relativistic particles which triggered a wave of papers of the subject.

The literature on quantum graphs is nowadays immense; we are not going to try to give a bibliographical review and refer instead to the proceedings volume of a recent Isaac Newton Institute programme [\refcite{ekkst08}] where one can find an extensive guide to further reading.

\subsection{Vertex coupling} \label{ss:coupling}

For simplicity let us consider a graph having a single vertex in form of a \emph{star}, i.e. $n$ halflines with the endpoint connected. The state Hilbert space $\mathcal{H}$  of such a system is $=\bigoplus_{j=1}^n L^2(\mathbb{R}_+)$ and the particle Hamiltonian acts on $\mathcal{H}$ as $\psi_j \mapsto -\psi''_j$; the values of physical constants are irrelevant for our discussion and we put conventionally $\hbar=2m=1$.

The Hamiltonian domain consists of $W^{1,2}$ functions; in order to make it self-adjoint we need to impose suitable boundary conditions at the vertex. Since we deal with a second-order operator, the latter involve boundary value
$\Psi(0):= \{\psi_j(0)\}$ and $\Psi'(0):= \{\psi'_j(0)\}$; conventionally they are written in the form
 \begin{equation} \label{bc:ks}
 A\Psi(0)+B\Psi'(0)=0
 \end{equation}
proposed by Kostrykin and Schrader [\refcite{ks99}], where the $n\times n$ matrices $A,B$ give rise to a self-adjoint operator \emph{iff} they satisfy the conditions
 \begin{itemize}
 \item $\;\mathrm{rank\,}(A|B)=n$
 \item $\;AB^*$ is self-adjoint
 \end{itemize}
The obvious drawback of \eqref{bc:ks} is that the pair $A,B$ is not unique. The common way to remove the non-uniqueness [\refcite{gg91, ha00, ks00}] is to choose
 \begin{equation} \label{bc:ha}
 A=U-I\,,\quad B=i(U+I)\,,
 \end{equation}
where $U$ is an $n\times n$ unitary matrix; there are also other unique forms more suitable for some purposes [\refcite{cet10a, cet10b, ku04}]; one of them we will need in~\sref{s:full} below. It is obvious from \eqref{bc:ha} that the coupling of $n$ edges is characterized in general by $n^2$ real parameters.

There is a simple way to derive the boundary conditions with which can be traced to [\refcite{ft00}] where it was used for $n=2$. Self-adjointness requires vanishing of the boundary form,
 $$ \sum_{j=1}^n (\bar\psi_j \psi'_j - \bar\psi'_j \psi_j)(0)=0\,, $$
which occurs \emph{iff} the norms $\|\Psi(0)\pm
i\ell\Psi'(0)\|_{\mathbb{C}^n}$ with a fixed $\ell\ne 0$ coincide, since the difference of the squared norms is just the \emph{lhs} of the displayed relation. Consequently, the vectors must be related by an $n\times n$ unitary matrix, which yields immediately $(U-I)\Psi(0)+i\ell(U+I)\Psi'(0)=0$. It may seem that we have an extra parameter here, however, matrices corresponding to two different values of $\ell$ are related by
 $$
 U' = \frac{(\ell+\ell')U +\ell-\ell'}{(\ell-\ell')U
 +\ell+\ell'}\,,
 $$
so it just fixes the length scale of the problem and we can put $\ell=1$ without loss of generality. Note also that the parameter matrix is closely related to the scattering at the vertex, specifically, it coincides with the on-shell scattering matrix at the momentum $k=1$.

\subsection{Examples of vertex coupling}

Denote by $\mathcal{J}$ the $n\times n$ matrix whose all entries are equal to one; then the unitary matrix $U= {2\over n+i\alpha} \mathcal{J}-I$ corresponds to the standard $\delta$ \emph{coupling} characterized by the conditions
 \begin{equation} \label{bc:delta}
 \psi_j(0)=\psi_k(0)=:\psi(0)\,,\; j,k=1,\dots,n\,,
 \quad
 \sum_{j=1}^n \psi'_j(0) = \alpha \psi(0)
 \end{equation}
of ``coupling strength'' $\alpha\in\mathbb{R}$; we include also the case $\alpha=\infty$, or $U=-I$, when the edges are decoupled with Dirichlet conditions at the endpoints. Another particular case of interest is $\alpha=0$ corresponding to the ``free motion''. It would be natural to call then \eqref{bc:delta} \emph{free} boundary conditions, however, they are mostly called \emph{Kirchhoff} in the literature\footnote{The name is generally accepted but unfortunate because in electricity it is associated with current conservation at the junction, and in the quantum case \emph{any} self-adjoint coupling preserves probability current. }. Note that the $\delta$-couplings are the only ones with wave functions continuous at the vertex.

The second example to mention is the $\delta'_s$ \emph{coupling}, a counterpart to the above one with the roles of functions and derivatives interchanged. The corresponding unitary matrix is $U= I-{2\over n-i\beta}\mathcal{J}$ giving
 \begin{equation} \label{bc:delta'}
  \psi'_j(0)=\psi'_k(0)=:\psi'(0)\,,\; j,k=1,\dots,n\,,
 \quad
 \sum_{j=1}^n \psi_j(0) = \beta \psi'(0)
 \end{equation}
with $\beta\in\mathbb{R}$; for $\beta=\infty$ we get decoupled edges with \emph{Neumann} conditions.

\section{Vertex understanding through approximations}

\subsection{Statement of the problem}

The first question to pose is why we should be interested in quantum graph vertex couplings. There are several reasons for that:
 \begin{itemize}
 \item One is mathematical. Different couplings define different Hamiltonians which have different spectral properties. Sometimes they can be quite involved; as an example let us number theoretic properties of rectangular lattice-graph spectra [\refcite{ex96a}].

 \item On a more practical side, the conductivity of nanostructures is controlled typically by application of external  fields. Understanding of vertex coupling would give us an alternative mean to this goal.

 \item As a specific example, the authors of Ref.~\refcite{ctf04} used the generalized point interaction on line as a model of a qubit; in a similar way star graphs with $n>2$ edges can similarly model \emph{qudits}.
 \end{itemize}
At a glance the vertex parameters can be interpreted easily. One should replace the graph in question by a family of ``fat graphs'', i.e. a tube network built around the graph skeleton, with appropriate Laplacian as the Hamiltonian. Such a system has no free parameters, so it would be enough to inspect the squeezing limit with the tube diameter tends to zero and to see which graph Hamiltonian we obtain. Unfortunately, as it is often the case with simple answers, the problem is in reality rather complicated:
 \begin{itemize}
 \item The answer depends substantially on the type of the Laplacian supported by the tube network. The \emph{Neumann} case is easier and after an effort more than a decade long an understanding was reached [\refcite{ep05, ep07, fw93, kz01, po06, rs01, sa00}]. The drawback was that the limit gave the free (Kirchhoff) boundary conditions only.
 \item the \emph{Dirichlet} case is more difficult and only recently some substantial results were obtained [\refcite{acf07, ce07, dc10, gr08, mv07, po05}], nevertheless, a lot of work remains to be done
 \end{itemize}
Before proceeding to our main topic, let us review briefly the existing results we have mentioned above.

\subsection{Briefly on Dirichlet networks}

The distinctive feature of the Dirichlet case is the energy blow-up associated with the fact that the transverse part of the Dirichlet Laplacian has lowest eigenvalue proportional to $d^{-2}$ where $d$ is the tube diameter. To get a meaningful result we have thus to use an \emph{energy renormalization} which can be done in different ways.
 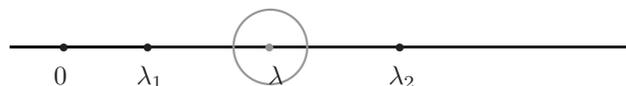
\begin{figure}
 \setlength\unitlength{1mm}
 \begin{picture}(95,0)(20,52)
 \thinlines
 \put(37.5,45){\line(1,0){80}}
 \thicklines
 \Black{\put(55.5,45){\line(1,0){65}}}
 \Black{\put(43.5,45){\circle*{1.2}}}
 \Black{\put(53.5,45){\circle*{1.2}}}
 \Gray{\put(68.5,45){\circle*{1.2}}}
 \Gray{\put(67.5,45){\circle{10}}}
 \Black{\put(83.5,45){\circle*{1.2}}}
 \put(37.5,40){\mbox{$0$}}
 \put(48.5,40){\mbox{$\lambda_1$}}
 \put(66,40){\mbox{$\lambda$}}
 \put(82,40){\mbox{$\lambda_2$}}
 \end{picture}
 \vspace{4em} \caption{Energy renormalization}
 \end{figure}
Molchanov and Vainberg [\refcite{mv07}] chose the energy $\lambda$ to be subtracted between the first and second transverse eigenvalue, cf.~Fig.~1, and obtained a nontrivial limit determined by scattering properties of the corresponding ``fat star''. A drawback of this approach is that leads to energy spectrum unbounded from below which is a feature one tries to avoid in meaningful nonrelativistic models.

Most authors choose therefore the transverse threshold $\lambda_1$ as the energy to subtract. In such a case the limit is \emph{generically trivial} giving disconnected edges with Dirichlet endpoints [\refcite{mv07, po05}]. However, the limit can be nontrivial provided the tube system we start with has a \emph{threshold resonance} [\refcite{acf07, ce07, gr08}]; a similar, closely related effect using finite star graphs was proposed in [\refcite{dc10}].

\subsection{A survey on Neumann network results}

Consider first for simplicity a finite connected graph $M_0$ with vertices $v_k$, $k \in K$ and edges $e_j \simeq I_j:=[0,\ell_j]$, $j \in J$; the corresponding Hilbert space is thus $L^2(M_0) := \bigoplus_{j\in J} L^2(I_j)$. The form $u\mapsto \|u'\|^2_{M_0}:= \sum_{j\in J} \|u'\|^2_{I_j}$
with $u\in W^{1,1}(M_0)$ is associated with the operator which acts as $-\Delta_{M_0}u = - u_j''$ \\ and satisfies the
\emph{free} boundary conditions.

\begin{figure}
\begin{center}
\includegraphics[height=3cm, width=9cm, angle=0]{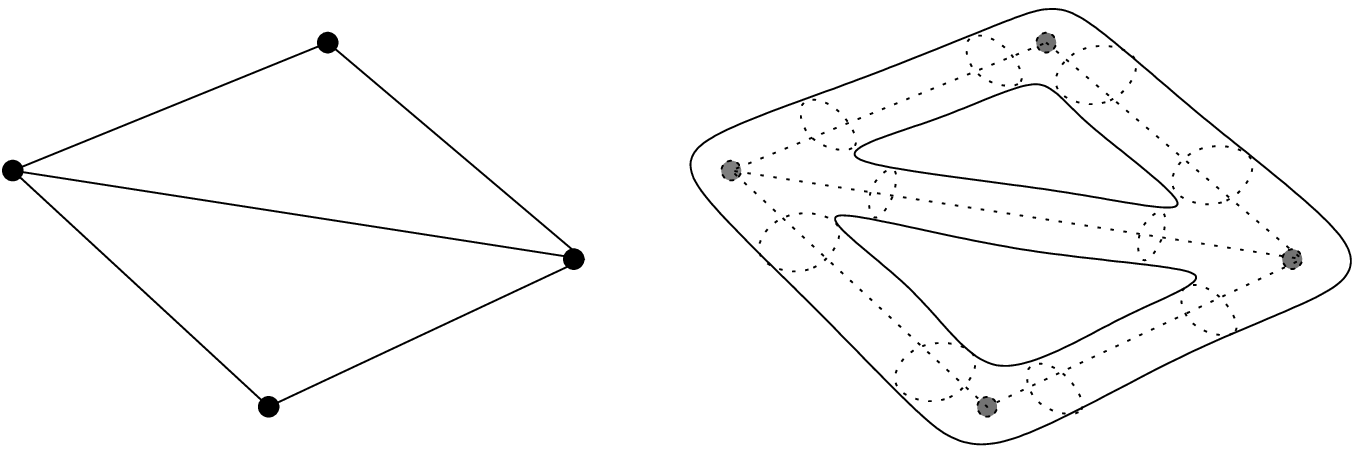}
\end{center}
\end{figure}
\vspace{-2em}
 \begin{figure}
 \begin{picture}(95,10)(30,16)
 \Black{
 \put(70,44){\mbox{$M_0$}}
 \put(200,44){\mbox{$M_\varepsilon$}}
 \put(135,57){\mbox{$e_j$}}
 \put(165,72){\mbox{$v_k$}}
 \put(275,52){\mbox{$U_{\varepsilon,j}$}}
 \put(305,65){\mbox{$V_{\varepsilon,k}$}} }
 \end{picture}
 \vspace{-2em} \caption{Graph $M_0$ and fat graph $M_\varepsilon$}
 \end{figure}
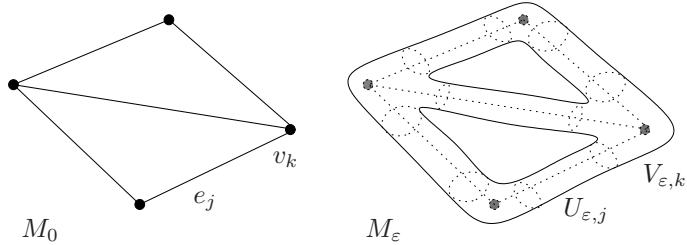

On the other hand, consider a Riemannian manifold $X$ of dimension $d \ge 2$ and the corresponding space $L^2(X)$ w.r.t. volume $\mathrm{d}X$ equal to $(\det g)^{1/2}\mathrm{d}x$ in a fixed chart. For $u \in C^\infty_\mathrm{comp}(X)$ we set
 \begin{equation} \label{lbLapl}
 q_X(u):=\|\mathrm{d}u\|^2_X = \int_{X}
 |\mathrm{d} u|^2 \mathrm{d}X\,,\quad |\mathrm{d} u |^2=
 \sum_{i,j} g^{ij} \partial_i u \, \partial_j
 \overline u\,;
 \end{equation}
the closure of this form is associated with the self-adjoint
\emph{Neumann} Laplacian $\Delta_X$ on the $X$. Let us stress that within this framework we can treat both ``solid'' tubes with the boundary at which Neumann condition is imposed, as well as ``sleeve-type'' manifolds without a boundary when the particle is supposed to live on the surface -- cf.~Fig.~2. This is made possible by the similarity of the transverse ground-state eigenfunction in both cases.

The ``fat graphs'' $M_\varepsilon$ associated with the graph $M_0$ are all constructed from $X$ by taking a suitable
\emph{$\varepsilon$-dependent family of metrics}. This the approach was used in [\refcite{ep05}]; in contrast to earlier work such as [\refcite{kz01}] one also need not assume that the network is embedded in a Euclidean space since only intrinsic geometrical properties are involved.

The analysis requires dissection of $M_\varepsilon$ into a union of edge and vertex components, $U_{\varepsilon,j}$ and
$V_{\varepsilon,k}$, respectively, with appropriate scaling properties,
 \begin{itemize}
 \item for edge regions we assume that $U_{\varepsilon,j}$ is diffeomorphic to $I_j \times F$ where $F$ is a compact and connected manifold (with or without a boundary) of dimension $m:=d-1$
 \item for vertex regions we assume that the manifold $V_{\varepsilon,k}$ is diffeomorphic to an $\varepsilon$-independent manifold $V_k$
\end{itemize}
In this setting one can prove the following result [\refcite{ep05}]:
 \begin{theorem}
Under the stated assumptions we have eigenvalue convergence, $\lambda_k(M_\varepsilon)\to \lambda_k(M_0)\,,\; k=1,2,\dots\,,$ as $\:\varepsilon\to 0$.
 \end{theorem}

The shrinking limit thus leads to free boundary conditions only, but also in other respects the stated result is not particularly strong, for instance, in that it concerns the eigenvalue convergence in finite graphs only. One can do better: in Ref.~\refcite{po06} Olaf Post proved a norm-resolvent convergence $\Delta_{M_\varepsilon}\to \Delta_{M_0}$ as $\varepsilon\to 0+$ on generally infinite graphs under natural uniformity conditions analogous to those of used in Theorem~\ref{gengraph}, namely (i) existence of nontrivial bounds on vertex degrees and volumes, edge lengths, and the second Neumann eigenvalues at vertices, (ii) appropriate scaling (analogous to the described above) of the metrics at the edges and vertices. The involved operators act on different Hilbert spaces, of course, and the stated limiting relation makes sense with a suitable identification map which we will describe below.

Other extensions are possible. For graphs with semi-infinite ``outer'' edges, e.g., the problem typically exhibits series of \emph{resonances}, and one may ask what happens with them if the graph is replaced by a family of ``fat'' graphs. Using \emph{exterior complex scaling} in the ``longitudinal'' variable one can prove a convergence result for
resonances as $\varepsilon\to 0\,$ [\refcite{ep07}]; the
same is true for \emph{embedded eigenvalues} of the
graph Laplacian which may remain embedded or become resonances for $\varepsilon>0$.

Hence we have a number of convergence results is available for squeezing limit of Neumann-type thin tube networks, however, the limiting operator corresponds always to \emph{free} boundary conditions only. The question is whether one can do better.

\section{Beyond the free coupling}

\subsection{A graph inspiration}

It is obvious that one has to add a new feature to the approximating family to get more general results. Let us look how one can approximate $\delta$ coupling on graphs using families of scaled potentials. For simplicity we will consider again the $n$-edge star graph as in \sref{ss:coupling}, however, we replace the Laplacian at the edges by a Schr\"odinger operator, $\psi_j \mapsto -\psi''_j+ V_j\psi_j$. In order make the problem well-defined we have to impose requirement on the potential; we suppose that $V_j\in L^1_\mathrm{loc}(\mathbb{R}_+)\,, \;j=1,\dots,n$. If the boundary conditions at the vertex are \eqref{bc:delta} with the parameter $\alpha\in\mathbb{R}$ we get a self-adjoint operator which we denote as $H_\alpha(V)$. Let now the potential contain a scaled component,
 \begin{equation} \label{squeeze}
 W_{\varepsilon,j}:= {1\over\varepsilon}\,
 W_j\left(x\over\varepsilon \right)\,,\quad j=1,\dots,n\,,
 \end{equation}
then we have the following result [\refcite{ex96b}]:
 \begin{theorem}
Suppose that the potentials $\:V_j \in L^1_\mathrm{loc} (\mathbb{R}_+)$ are below bounded and $\:W_j\in L^1(\mathbb{R}_+)$ for $\:j=1,\dots,n\,$. Then
 $$
 H_0(V+W_{\varepsilon})\longrightarrow H_{\alpha}(V)
 $$
as $\varepsilon\to 0+$ in the norm resolvent sense, with the
coupling parameter defined as $\alpha:= \sum_{j=1}^n \int_0^{\infty} W_j(x)\,dx$.
 \end{theorem}
Our aim is to ``lift'' this result to tube networks.

\subsection{Single vertex networks}

Consider first a star graph again. Let $G$ have one vertex $v$ and $\mathrm{deg}\,v$ adjacent edges of lengths $\ell_e
\in (0,\infty]$. The corresponding Hilbert space is $L^2(G):= \bigoplus_{e \in E} L^2(I_e)$, the \emph{decoupled} Sobolev
space of order $k$ is $W^{2,k}_\mathrm{max}(G) :=
\bigoplus_{e \in E} W^{2,k}(I_e)$ together with its natural norm.

Let $\underline{p}=\{p_e\}$ have components $p_e>0$ for $e \in E$; we introduce it because we want to consider squeezing limits also in the situation when the tubes have different cross sections. The Sobolev space associated with weight $\underline{p}$ is
$$
  W^{2,k}_{\underline{p}}(G)
     := \left\{f \in W^{2,k}_{\mathrm{max}}(G):\: \underline{f} \in \mathbb{C} \underline{p} \right\}\,,
$$
where $\underline{f} := \{f_e(0)\}$, in particular, if all the components are equal, $\underline{p}=(1,\dots,1)$, we arrive at the \emph{continuous}
Sobolev space $W^{2,1}(G) := W^{2,1}_{\underline{p}}(G)$.

Next we have to introduce operators on the graph. We start with the (weighted) \emph{free} Hamiltonian $\Delta_G$ defined via the quadratic form $\mathfrak{d}=\mathfrak{d}_G$,
$$
  \mathfrak{d}(f) := \|f'\|^2_G = \sum_e
  \|e'\|^2_{I_e}
  \quad\mathrm{and}\quad
  \mathrm{dom}\,\mathfrak{d} := W^{1,1}_{\underline{p}}(G)
$$
for a fixed $\underline{p}\:$ (we drop the index $\underline{p}$); the form is a closed as related to the Sobolev norm $\|f\|^2_{W^{1,1}(G)} = \|f'\|^2_G + \|f\|^2_G$.
The Hamiltonian with \emph{$\delta$-coupling of
strength $q$} is defined via the quadratic form $\mathfrak{h}=\mathfrak{h}_{(G,q)}$ given by
$$
  \mathfrak{h}(f) := \|f'\|^2_G + q(v) |f(v)|^2 \quad\mathrm{and}\quad
  \mathrm{dom}\, \mathfrak{h} := W^{2,1}_{\underline{p}}(G)
$$
Using standard Sobolev arguments one can show [\refcite{ep09}] that the $\delta$-coupling is a ``small'' perturbation of $\Delta_G$ by estimating the difference $\mathfrak{h}(f) - \mathfrak{d}(f)$.

The manifold model of the ``fat'' graph is constructed as in the previous section. Given $\varepsilon \in (0, \varepsilon_0]$ we associate a $d$-dimensional manifold $X_\varepsilon$ to the graph $G$ in the following way: to the edge $e \in E$ and the vertex $v$ we ascribe the Riemannian manifolds
$$
  X_{\varepsilon,e} := I_e \times \varepsilon Y_e \quad\mathrm{and}\quad
  X_{\varepsilon,v} := \varepsilon X_v\,,
$$
respectively, where $\varepsilon Y_e$ is a manifold $Y_e$ equipped with metric $h_{\varepsilon,e}:=\varepsilon^2 h_e$ and $\varepsilon X_{\varepsilon,v}$ carries the metric $g_{\varepsilon,v}=\varepsilon^2 g_v\,$. As before, we use the $\varepsilon$-independent coordinates $(s,y) \in
X_e=I_e \times Y_e$ and $x \in X_v$, so the radius-type parameter $\varepsilon$ only enters via the Riemannian metric.
Let us stress this includes the case of the $\varepsilon$-neighborhood of an embedded graph $G \subset \mathbb{R}^d$, but only  up to a longitudinal error of order of $\varepsilon$; this problem can be dealt with again using an $\varepsilon$-dependence of the metric in the longitudinal
direction.

The Hilbert space of the manifold model $L^2(X_\varepsilon)$ can be decomposed as
$$
  L^2(X_\varepsilon)
     = \bigoplus_e \big( L^2(I_e) \otimes L^2(\varepsilon Y_e)\big) \oplus L^2(\varepsilon X_v)
$$
with the norm given accordingly by
$$
  \|u\|^2_{X_\varepsilon}
  = \sum_{e \in E} \varepsilon^{d-1} \int_{X_e} |u|^2 \mathrm{d} y_e \mathrm{d} s
   + \varepsilon^d \int_{X_v} |u|^2 \mathrm{d} x_v\,,
$$
where $\mathrm{d} x_e=\mathrm{d} y_e \mathrm{d} s$ and $\mathrm{d} x_v$ denote the Riemannian volume measures associated to the (unscaled) manifolds $X_e=I_e
\times Y_e$ and $X_v$, respectively. We also introduce  the Sobolev space $W^{1,1}(X_\varepsilon)$ of order one defined conventionally as the completion of the space of smooth functions with compact support under the norm $\|u\|^2_{W^{1,1} (X_\varepsilon)} = \|\mathrm{d}u\|^2_{X_\varepsilon} + \|u\|^2_{X_\varepsilon}$.

Next we pass to operators on the manifold. The Laplacian $\Delta_{X_\varepsilon}$ on $X_\varepsilon$ is given via its
quadratic form $\mathfrak{d}_\varepsilon (u)$ equal to
$$
  \|\mathrm{d}u\|^2_{X_\varepsilon}
  = \sum_{e \in E} \varepsilon^{d-1} \int_{X_e}
          \Bigl(|u'(s,y)|^2
                + \frac 1 {\varepsilon^2} |\mathrm{d}_{Y_e} u|^2_{h_e}
          \Bigr) \mathrm{d} y_e \mathrm{d} s
   + \varepsilon^{d-2} \int_{X_v} |\mathrm{d} u|_{g_v}^2\, \mathrm{d} x_v
$$
where $u'$ is the \emph{longitudinal} derivative, $u'=\partial_s u$, and $\mathrm{d} u$ is the exterior derivative of $u$. Again, $\mathfrak{d}_\varepsilon$ is closed by definition. Adding a potential, we define the Hamiltonian $H_\varepsilon$ as the operator associated with the form $\mathfrak{h}_\varepsilon = \mathfrak{h}_{(X_\varepsilon, Q_\varepsilon)}$ given by
$$
  \mathfrak{h}_\varepsilon = \|\mathrm{d}u\|^2_{X_\varepsilon}
  + \langle u, Q_\varepsilon u\rangle_{X_\varepsilon}\,,
$$
where the potential $Q_\varepsilon$ is supported in the vertex region $X_v$ only. Now we use graph result mentioned as an inspiration and choose 
$$
  Q_\varepsilon(x) = \frac 1 \varepsilon Q(x)\,,
$$
where $Q=Q_1$ is a fixed bounded and measurable function on $X_v$. The reader may wonder that in comparison to \eqref{squeeze} the factor $\varepsilon^{-1}$ is missing in the argument, however, this is due to our choice to perform the squeezing by the change of the metric only.

One can establish the relative (form-)boundedness of $H_\varepsilon$ with respect to the free operator $\Delta_{X_\varepsilon}\,$: to a given $\eta \in (0,1)$ there is $\varepsilon_\eta>0$ such that the form $\mathfrak{h}_\varepsilon$ is relatively form-bounded with respect to the free form $\mathfrak{d}_\varepsilon$, that is,
there is $\widetilde{C}_\eta>0$ such that
  $$
    |\mathfrak{h}_\varepsilon(u) - \mathfrak{d}_\varepsilon(u)|
    \le \eta \, \mathfrak{d}_\varepsilon(u) + \widetilde{C}_\eta \|u\|^2_{X_\varepsilon}
  $$
whenever $0 < \varepsilon\le \varepsilon_\eta$ with explicit constants $\varepsilon_\eta$ and $\widetilde{C}_\eta$. The latter are given explicitly in [\refcite{ep09}]; what is important that they are expressed in terms of the parameters of the model which we give below.

We have mentioned above that our operators acts in different spaces, namely the Hilbert spaces $\mathcal{H}=L^2(G)$ and $\widetilde{\mathcal{H}_\varepsilon}=L^2(X_\varepsilon)$ and their Sobolev counterparts, hence we need to define quasi-unitary operators to relate the graph and manifold Hamiltonians. For further purposes we denote
$$
  p_e:= (\mathrm{vol}_{d-1} Y_e)^{1/2} \quad\mathrm{and}\quad
  q(v) = \int_{X_v} Q \mathrm{d} x_v\,;
$$
recall that we introduced the weights $p_e$ to be able to treat situations when the tube cross sections $Y_e$ are mutually different.

First we define the graph-to-manifold map, $J : \mathcal{H} \to \widetilde{\mathcal{H}_\varepsilon}$, by
 \begin{equation} \label{ident}
  J f := \varepsilon^{-{(d-1)/2}}\bigoplus_{e \in E} (f_e \otimes {1\!\!\!\!-}_e) \oplus 0\,,
 \end{equation}
where ${1\!\!\!\!-}_e$ is the normalized eigenfunction of $Y_e$
associated to the lowest (zero) eigenvalue, i.e.
${1\!\!\!\!-}_e(y)=p_e^{-1}$. Next introduce the following averaging operators
$$
  {\textstyle{\int\!\!\!\!\!-}}_v u := {\int\!\!\!\!\!\!-}_{X_v} u \mathrm{d} x_v
           \quad\mathrm{and}\quad
  {\textstyle{\int\!\!\!\!\!-}}_e u(s) := {\int\!\!\!\!\!\!-}_{Y_e} u(s,\cdot) \mathrm{d} y_e
$$
The map in the opposite direction, $J' : \widetilde{\mathcal{H}}_\varepsilon \to \mathcal{H}$, is given by the adjoint,
$$
  (J' u)_e (s) = \varepsilon^{(d-1)/2} \langle {1\!\!\!\!-}_e, u_e(s,\cdot) \rangle_{Y_e}
  = \varepsilon^{(d-1)/2} p_e {\textstyle{\int\!\!\!\!\!-}}_e u(s)\,.
$$
In an analogous way one can construct identification maps between the Sobolev spaces. They are need in the proofs but not for stating the result, hence we refer the reader to [\refcite{ep09}] for their explicit forms.

Using these notions in combination with an abstract convergence result of [\refcite{po06}] one can then arrive at the following conclusions [\refcite{ep09}]:

 \begin{theorem}
As $\varepsilon \to 0$, we have
  \begin{eqnarray*}
      \|J(H - z)^{-1} - (H_\varepsilon - z)^{-1}J\|
      &=& \mathcal{O}(\varepsilon^{1/2}),\\
      \|J(H - z)^{-1}J' - (H_\varepsilon - z)^{-1}\| &=&
      \mathcal{O}(\varepsilon^{1/2})
  \end{eqnarray*}
for $z \notin [\lambda_0,\infty)$. Moreover, $\phi(\lambda) =(\lambda-z)^{-1}$ can be replaced by any measurable, bounded function converging to a constant as $\lambda \to \infty$ and being continuous in a neighborhood of $\sigma(H)$.
 \end{theorem}

 \begin{corollary}
 The spectrum of $H_\varepsilon$ converges to $\sigma(H)$ uniformly on any finite energy interval, and the same is true for the essential spectra.
 \end{corollary}

 \begin{corollary}
 For any $\lambda \in \sigma_\mathrm{disc}(H)$ there exists a family $\{\lambda_\varepsilon\}$ with $\lambda_\varepsilon \in \mathrm{disc}(H_\varepsilon)$ such that $\lambda_\varepsilon \to \lambda$ as $\varepsilon \to 0$, and moreover, the multiplicity is preserved. If $\lambda$ is a simple eigenvalue with normalized eigenfunction $\phi$, then there exists a family of simple normalized eigenfunctions $\{\phi_\varepsilon\}_\varepsilon$ of $H_\varepsilon$ such that $\|J\phi - \phi_\varepsilon\|_{X_\varepsilon} \to 0$ holds as $\varepsilon \to 0$.
 \end{corollary}

\subsection{The general case}

So far we have talked for simplicity about the star-shaped graphs only. The same technique of ``cutting'' the graph and the corresponding manifold into edge and vertex regions works also in the general case. As a result of the analysis performed in Ref.~\refcite{ep09} we get

 \begin{theorem} \label{gengraph}
 Assume that $G$ is a metric graph and $X_\varepsilon$ the corresponding approximating manifold. If
  $$
    \sup_{v \in V} \frac{\mathrm{vol}\, X_v}{\mathrm{vol}\, \partial X_v} < \infty\,,\quad
    \sup_{v \in V} \|Q \!\upharpoonright\!{X_v}\|_\infty < \infty\,,\quad
    \inf_{e \in E} \ell_e > 0
  $$
and
  $$
    \inf_{v \in V} \lambda_2(v) > 0\,,\qquad
    \inf_{e \in E} \lambda_2(e) > 0\,,
  $$
where $\lambda_2$ denotes the \emph{second} Neumann eigenvalue in the appropriate manifold region,  then the corresponding Hamiltonians $H=\Delta_G + \sum_v q(v) \delta_v$ and $H_\varepsilon = \Delta_{X_\varepsilon} + \sum_v \varepsilon^{-1} Q_v$ are $\mathcal{O} (\varepsilon^{1/2})$-close with the error depending only on the above indicated global constants.
 \end{theorem}
In this way we have managed to solve the problem for quantum graphs with $\delta$-couplings under mild uniformity conditions.

\section{Discontinuity at the vertex: the example of $\delta'_s$}

While the above results break the Kirchhoff restriction of the previous studies, they do not give a full answer; recall that the $\delta$-couplings at a vertex $v$ represent a one-parameter subfamily in the $n^2$ parameter family, $n=\mathrm{deg}\,v$, of all self-adjoint couplings. Let us now investigate the case of $\delta'_s$ as a prime example of coupling with functions discontinuous at the vertex.

\subsection{The idea of Cheon and Shigehara}

Our strategy will be the same as before, first we will construct an approximation on the graph itself and then we we will lift it to the manifold. The problem is not easy and its core is the question whether one can approximate the $\delta'$-interaction on the line by means of (regular or singular) potentials. It was believed for a considerable time that this problem has no solution, until Cheon and Shigehara in the seminal paper [\refcite{cs98}] demonstrated a formal approximation by means of of three $\delta$-interaction; a subsequent mathematical analysis [\refcite{an00, enz01}] showed that it converges in fact in norm-resolvent sense.

 \begin{figure}
 \setlength\unitlength{1mm}
 \begin{picture}(95,0)(45,52)
 \thicklines
 \Black{
 \put(70,40){\line(-1,-1){10}}
 \put(70,40){\line(2,-1){12}}
 \put(70,40){\line(1,1){15}}
 \put(70,40){\line(-1,1){10}}}
 \Gray{\put(68.7,40){\circle*{.8}}
 \thinlines
 \put(68.7,40){\line(0,-1){18}}
 \put(65.5,36.7){\line(0,-1){8}}
 \put(72.2,38.4){\line(0,-1){8}}
 \put(71.4,42.6){\line(0,-1){8}}
 \put(66,42.7){\line(0,-1){8}} }
 \thicklines
 \Black{
 \put(120,40){\line(-1,-1){10}}
 \put(120,40){\line(2,-1){12}}
 \put(120,40){\line(1,1){15}}
 \put(120,40){\line(-1,1){10}}}
 \Gray{\put(118.9,40){\circle*{1.0}}}
 \put(89.5,39){\mbox{\Large $\longrightarrow$}}
 \put(90,35){\mbox{$a\to 0$}}
 \put(118,44){\mbox{$\beta$}}
 \put(66,42.5){\mbox{$a$}}
 \put(68,24){\mbox{$b(a)$}}
 \put(71,30.8){\mbox{$c(a)$}}
 \Black{\put(103,52){\mbox{$H_\beta$}}
 \put(51,52){\mbox{$H^{b,c}$}}}
 \end{picture}
 \vspace{8em} \caption{CS approximation scheme on a graph}
 \end{figure}
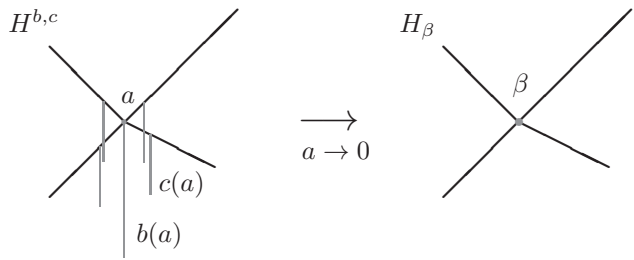

The idea can be extended to $\delta'_s$-coupling on a graph. A scheme of the approximation is given on Fig.~3. One starts with a $\delta$-coupling of strength $b(a)$ and adds $\delta$-interactions of strength $c(a)$ at the graph edges; the parameter $a$ is the distance of the additional interactions from the vertex. Core of the approximation lies in a suitable, $a$-dependent choice of the interaction strengths: we put
$$
   H^{\beta,a} := \Delta_G + b(a) \delta_{v_0} + \sum_e c(a) \delta_{v_e},
  \quad
  b(a) = - \frac \beta{a^2}, \quad
  c(a) = - \frac 1a
$$
which corresponds to the quadratic form
$$
   {\mathfrak{h}}^{\beta,a}(f) := \sum_e \|f_e'\|^2
     - \frac \beta{a^2} |f(0)|^2 - \frac 1 a \sum_e |f_e(a)|^2,
  \quad
  \mathrm{dom}\,  \mathfrak{h}^a = W^{2,1}(G)\,.
$$
Then we have the following result [\refcite{ce04}]:
 \begin{theorem} \label{t:ce04}
 $\|( H^{\beta,a} - z)^{-1} - (H^\beta - z)^{-1}\| =\mathcal{O}(a)$ holds as $a\to 0$ for $z \notin \mathbb{R}$, where $\|\cdot\|$ is the operator norm on $L^2(G)$.
 \end{theorem}
\emph{Proof} is by a direct computation. We note that the result is highly non-generic, both resolvents are strongly singular as $a\to 0$ but in the difference those singularities cancel.

\subsection{The convergence result}

\begin{figure}
\begin{picture}(0,0)%
\includegraphics[scale=0.8]{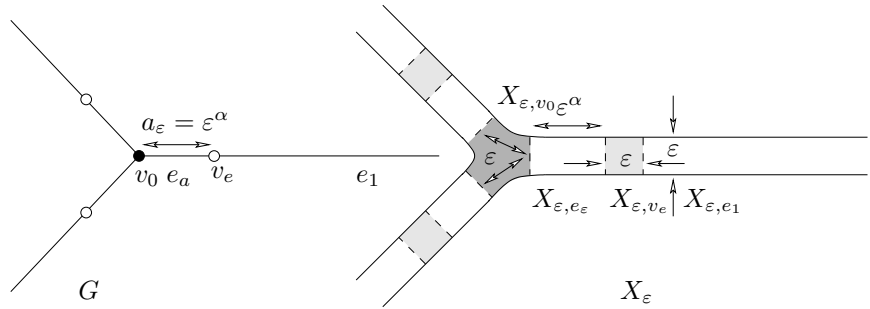}%
\end{picture}%
\setlength{\unitlength}{3315sp}%
\begin{picture}(6431,2281)(79,-1565)
\put(4650,-1496){$X_\varepsilon$}%
\put(598,-1496){$G$}%
\put(1070,-227){$a_\varepsilon=\varepsilon^\alpha$}%
\put(1021,-622){$v_0$}%
\put(1586,-594){$v_e$}%
\put(1256,-622){$e_a$}%
\put(2676,-622){$e_1$}%
\put(5000,-424){$\varepsilon$}%
\put(4171,-114){$\varepsilon^\alpha$}%
\put(3635,-481){$\varepsilon$}%
\put(4650,-500){$\varepsilon$}%
\put(4569,-807){$X_{\varepsilon,v_e}$}%
\put(3986,-807){$X_{\varepsilon,e_\varepsilon}$}%
\put(5120,-807){$X_{\varepsilon,e_1}$}%
\put(3731,-033){$X_{\varepsilon,v_0}$}%
\end{picture}%
  \caption{Scheme of the lifting}
\end{figure}

\noindent Now we will lift the above graph approximation result to the manifold according to the scheme depicted on Fig.~4. For simplicity assume that the star graph in question is finite with all edges having the same length; without loss of generality we may put it equal to one. In contrast to the previous section we have two parameters to deal with, the tube width $\varepsilon$ and the distance of the additional potentials; we choose $a=a_\varepsilon=\varepsilon^\alpha$ with $\alpha\in(0,1)$ to be specified later. The crucial point is the choice of the additional potentials. The simplest option is to assume that they are constant,
 $$
  Q_{\varepsilon,v} (x) := \frac 1 \varepsilon \cdot
           \frac {q_\varepsilon (v)}{\mathrm{vol}\, X_v}, \qquad x \in X_v
 $$
so that $\int_{X_v} Q_{\varepsilon,v} \mathrm{d} x = \varepsilon^{-1}q_\varepsilon(v)$, where we put
 $$
  q_\varepsilon(v_0) := b(\varepsilon^\alpha)
    = - \beta \varepsilon^{-2\alpha}
    \quad \mathrm{and} \quad
  q_\varepsilon(v_e) := c(\varepsilon^\alpha)
    = - \varepsilon^{-\alpha}\,.
 $$
The corresponding manifold Hamiltonian and the respective
quadratic form are then given by
 $$
  \label{eq:h.beta.eps}
  H_\varepsilon^\beta
  = \Delta_{X_\varepsilon}
  - \varepsilon ^{-1-2\alpha} \frac \beta {\mathrm{vol}\, X_{v_0}} \chi_{X_{v_0}}
  - \varepsilon^{-1-\alpha} \sum_{e \in E} \chi_{X_{v_e}}\,,
 $$
where $\chi_X$ is the indicator function of the set $X$, and
 $$
  \mathfrak{h}_\varepsilon^\beta(u)
  = \|\mathrm{d}u\|^2_{X_\varepsilon}
  - \varepsilon ^{-1-2\alpha} \frac \beta {\mathrm{vol}\, X_{v_0}} \|u\|^2_{X_{\varepsilon,v_0}}
  - \varepsilon^{-1-\alpha} \sum_{e \in E} \|u\|^2_{X_{\varepsilon,v_e}}\,,
 $$
respectively. Note that the unscaled vertex neighborhood $X_{v_e}$ of each of the added vertices $v_e$ has volume one by construction.

We employ again the identification operator \eqref{ident}. Using the same technique as in the $\delta$ case, one can prove the following result [\refcite{ep09}]:

 \begin{theorem} \label{t:delta'}
 Assume that $0 < \alpha < 1/13$, then
  $$
    \left\|(H_\varepsilon^\beta-i)^{-1} J - J
      (H^\beta -i)^{-1} \right\| \to 0
  $$
 holds as the radius parameter $\varepsilon \to 0$.
 \end{theorem}

 \begin{remark}
The theorem has analogous corollaries as the $\delta$-coupling result of the previous section, however, a caveat is due. If $\beta<0$ the the spectrum of $H^{\beta,a}$ is uniformly bounded from below as $a\to 0$. If $\beta \ge 0$, on the other hand, the spectrum of $ H^{\beta,a}$ is asymptotically unbounded from below, $\inf \sigma(H^{\beta,a}) \to -\infty$ as  $a \to 0$. At the same time, for $\beta \ge 0$ the spectrum of the approximating operator $H_\varepsilon^\beta$ is asymptotically unbounded from below, $\inf \sigma(H_\varepsilon^\beta)\to -\infty$ as $\varepsilon \to 0$. This fact, existence of eigenvalues which escape to $-\infty$ in the limit does not contradict the fact that the limit operator $H^\beta$ is non-negative. Recall that the spectral convergence holds only for \emph{compact} intervals $I \subset \mathbb{R}$, in particular, $\sigma(H^\beta) \cap I = \emptyset$ implies that $\sigma(H_\varepsilon^\beta) \cap I = \emptyset$ and $\sigma{H^{\beta,\varepsilon}} \cap I = \emptyset$ for $\varepsilon>0$ small enough.
 \end{remark}

 \begin{remark}
While it is easy to see that the parameter $\alpha$ in the approximation must less than one, the value $\frac{1}{13}$ is certainly not optimal.
 \end{remark}

\section{Full solution on the graph level} \label{s:full}

\subsection{Going beyond $\delta$ and $\delta'_s$}

The network approximations of the $\delta$ and $\delta'_s$-couplings described in the two previous sections represent the present state of art in this question. One naturally asks whether one can extend the technique to other vertex couplings. Following the philosophy used here we should look first whether such approximations exist on the graph level.

The simplest extension covers the class of couplings invariant w.r.t. permutations of edges. It is a two-parameter family containing $\delta$ and $\delta'_s$ as particular cases; in the parametrization \eqref{bc:ha} its elements are characterized by matrices $U=a\mathcal{J}+bI$ with $|b|=1$ and $|b +a\,\mathrm{deg}\,v|=1$. The appropriate approximation in the spirit of Theorem~\ref{t:ce04} was worked out in Ref.~\refcite{et06}; note that, as with $\delta$ and $\delta'_s$, the problem again splits into a one-dimensional component in the subspace symmetric over the edges and its complement which is trivial from the coupling point of view.

If we relax the symmetry requirement things become more complicated. The first question is what we can achieve by modifications of the original Cheon and Shigehara idea, placing a finite number of properly scaled $\delta$-interactions on each edge. The answer is given by the following claim [\refcite{et07}]:

 \begin{proposition}
 Let $G$ be an $n$-edged star graph and $G(d)$ obtained by adding a finite number of $\delta$s at each edge, uniformly in $d$, at the distances $\mathcal{O}(d)$ as $d\to 0_+$. Suppose that these approximations yield conditions \eqref{bc:ks} with some $A,\,B$ as $d\to 0$. The family which can be obtained in this way depends on $2n$ parameters if $n>2$, and on three parameters for $n=2$.
 \end{proposition}

\noindent It was demonstrated in Ref.~\refcite{et07} that a family with the maximum number of parameters given in the proposition can be indeed constructed.

In order to get a wider class one has to pass to a more general approximation. The idea put forward in Ref.~\refcite{et07} was \emph{to change locally the graph topology} by adding new edges in the vicinity of the  vertex whose lengths shrink to zero in the approximation. This yielded a family of couplings with ${n+1\choose 2}$ parameters and real matrices $A,B$. To get a better result which will be described below one has to do two more things:
 \begin{itemize}
 \item together with adding edges in the vicinity of the vertex one has also to \emph{remove parts of the graph} to optimize locally the approximating graph topology,
 \item furthermore, one has to add \emph{local magnetic fields} described by suitable vector potentials to be able to get couplings which are not invariant w.r.t. time reversion.
 \end{itemize}

\subsection{An alternative unique parametrization}

In order to present the indicated approximation result we have to first introduce another form of the boundary conditions \eqref{bc:ks} derived in Ref.~\refcite{cet10a}.

 \begin{theorem}
 Consider a quantum graph vertex of degree $n$. If $m\leq n$, $S\in\mathbb{C}^{m,m}$ is a self-adjoint matrix and $T\in\mathbb{C}^{m,n-m}$, then the relation
 \begin{equation}\label{bc:st}
 \left(\begin{array}{cc}
 I^{(m)} & T \\
 0 & 0
 \end{array}\right)\Psi'=
 \left(\begin{array}{cc}
 S & 0 \\
 -T^* & I^{(n-m)}
 \end{array}\right)\Psi
 \end{equation}
expresses self-adjoint boundary conditions of the type \eqref{bc:ks}. Conversely, for any self-adjoint vertex coupling there is an $m\leq n$ and a numbering of the edges such that the coupling is described by the boundary conditions \eqref{bc:st} with uniquely given matrices $T\in\mathbb{C}^{m,n-m}$ and self-adjoint $S\in\mathbb{C}^{m,m}$.
 \end{theorem}

 \begin{remark}
As we have mentioned there are several unique forms of the conditions \eqref{bc:ks}. Kuchment [\refcite{ku04}] splits the boundary value space using projections $P,Q$ corresponding to Dirichlet, $P\Psi=0$, and Neumann, $Q\Psi'=0$, parts and the mixed conditions in the complement. It is easy to see that parts singled out correspond to eigenspaces of $U$ corresponding to eigenvalues $\mp 1$, respectively. The conditions \eqref{bc:st} which one call the \emph{ST-form} single out the eigenspace corresponding to $-1$. There is also an analogue of \eqref{bc:st} symmetric w.r.t. the two singular parts, called \emph{PQRS-form}, cf. Ref.~\refcite{cet10b}.
 \end{remark}

\subsection{A general graph approximation}

In view of the above result one can put general self-adjoint boundary conditions into the form \eqref{bc:st} renumbering the edges if necessary. We will now describe how those can be approximated by a family of graphs with locally changed topology and added magnetic fields. For notational purposes we adopt the following convention:  the lines of the matrix $T$ are indexed from 1 to $m$, the columns are indexed from $m+1$ to $n$.

The general vertex-coupling approximation, schematically depicted in Fig.~5, consists of the following sequence of steps:

\begin{figure}[!t]
\begin{center}
\includegraphics[width=5cm, keepaspectratio]{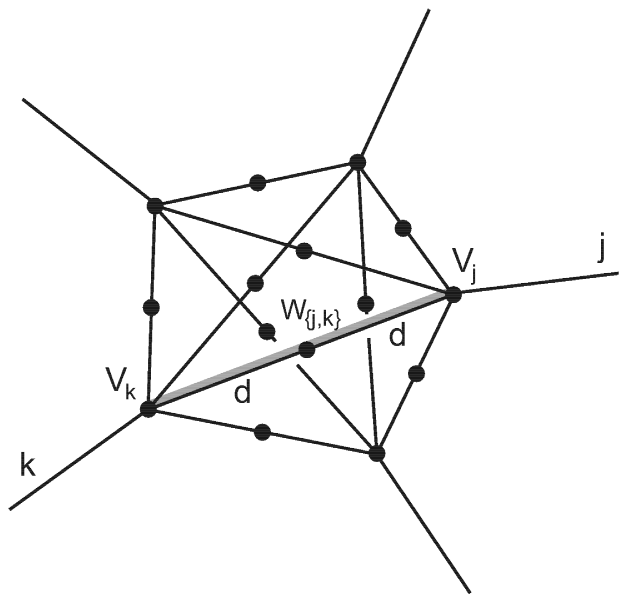}
\end{center}
\caption{The scheme of the approximation. All the inner links are of length $2d$, some may be missing. The grey line symbolizes the vector potential $A_{(j,k)}(d)$.}
\end{figure}

\begin{itemize}
 \item Take $n$ halflines, each parametrized by $x\in\mathbb{R}_+$, with the endpoints denoted as $V_j$, and put a $\delta$-coupling to the edges specified below with the parameter $v_j(d)$ at the point $V_j$ for all $j=1,\dots,n$.

 \item Some pairs $V_j,V_k,\; j\neq k$, of halfline endpoints are connected by edges of length $2d$, and the center of each such joining segment is denoted as $W_{\{j,k\}}$. This happens if one of the following conditions is satisfied:
 \begin{itemize}
 \item[(a)] $j=1,\dots,m$, $k\geq m+1$, and $T_{jk}\neq0$ (or $j\geq m+1$, \\ $k=1,\dots,m$, and $T_{kj}\neq0$),
 \item[(b)] $j,k=1,\dots,m$, and $\,S_{jk}\neq0$ or $\;(\,\exists l\geq m+1\,)\;\\ (\,T_{jl}\neq0\wedge T_{kl}\neq0\,)$.
\end{itemize}

 \item At each middle-segment point $W_{\{j,k\}}$ we place a $\delta$ interaction with a parameter $w_{\{j,k\}}(d)$. The connecting edges of length $2d$ are considered as consisting of two segments of length $d$, and on each of them the variable runs from zero at  $W_{\{j,k\}}$ to $d$ at the points $V_j,V_k$.

 \item On each connecting segment we put a vector potential of constant value between the points $V_j$ and $V_k$. We denote its strength between the points $W_{\{j,k\}}$ and $V_j$ as $A_{(j,k)}(d)$, and between the points $W_{\{j,k\}}$ and $V_k$ as $A_{(k,j)}(d)$. It follows from the continuity that $A_{(k,j)}(d)=-A_{(j,k)}(d)$ for any pair $\{j,k\}$.

\end{itemize}
The choice of the dependence of $v_j(d)$, $w_{\{j,k\}}(d)$ and
$A_{(j,k)}(d)$ on the parameter $d$ is crucial for the
approximation. In order to describe it we introduce the set $N_j\subset\{1,\dots,n\}$ containing indices of all the edges that are joined to the $j$-th one by a connecting segment, i.e.
 \begin{align}\label{Nj}
 N_j=&\{k\le m|\, S_{jk}\neq0\}\cup\{k\le m|\,
  (\exists l\geq m+1)(T_{jl}\neq0\wedge T_{kl}\neq0)\} \nonumber \\
  &\cup\{k\geq m+1|\, T_{jk}\neq0\} \qquad \text{for } j\le m\\ N_j=&\{k\le m|\, T_{kj}\neq0\} \qquad\qquad\quad \text{for } j\geq m+1 \nonumber
 \end{align}
We distinguish two cases regarding the indices involved:

\smallskip

\noindent \textit{Case I.} First we suppose that
$j=1,\dots,m$ and $l\in N_j\backslash\{1,\dots,m\}$. Then the vector potential strength may be chosen as follows,
 $$
A_{(j,l)}(d)=\left\{\begin{array}{lcl} \frac{1}{2d}\arg\,T_{jl} &
\;\text{if} & \;\mathrm{Re}\, T_{jl}\geq0\,, \\
[.5em] \frac{1}{2d}\left(\arg\,T_{jl}-\pi\right) & \;\text{if} &
\;\mathrm{Re}\, T_{jl}<0
\end{array}\right.
 $$
while for $v_l$ and $w_{\{j,l\}}$ with $l\geq m+1$ we put
 $$
v_l(d)=\frac{1-\#N_l+\sum_{h=1}^m\langle T_{hl}\rangle}{d}
\qquad \forall l\geq m+1\,,
 $$
 $$
w_{\{j,l\}}(d)=\frac{1}{d}\left(-2+\frac{1}{\langle T_{jl}\rangle}\right)
\qquad \forall j,l\quad\text{indicated above}\,,
 $$
where the symbol $\langle\cdot\rangle$ here has the following meaning: if $c\in\mathbb{C}$, then
 $$
\langle c\rangle=\left\{\begin{array}{ccl}
|c| & \text{if} & \mathrm{Re}\, c\geq0\,, \\
-|c| & \text{if} & \mathrm{Re}\, c<0\,.
\end{array}\right.
 $$
We remark that the choice of $v_l(d)$ is not unique. This is related to the fact that for $m=\mathrm{rank}\,B<n$ the number of parameters of the coupling is reduced from $n^2$ to at most $n^2-(n-m)^2$.

\smallskip

\noindent \textit{Case II.} Suppose next that $j=1,\dots,m$ and $k\in N_j\cap\{1,\dots,m\}$.

 $$
A_{(j,k)}(d)=\frac{1}{2d}\arg\,\left(d\cdot S_{jk}
+\sum_{l=m+1}^n T_{jl}\overline{T_{kl}} - \mu\pi \right)\,,
 $$
where $\mu=0$ if
 $$
\mathrm{Re}\left(d\cdot S_{jk}
+\sum_{l=m+1}^n T_{jl}\overline{T_{kl}}\right)\ge 0
 $$
and $\mu=1$ otherwise. The functions $w_{\{j,k\}}$ are given by
 $$
w_{\{j,k\}} = - \frac 1d \left( 2 +
\left\langle d\cdot S_{jk}
+\sum_{l=m+1}^n T_{jl}\overline{T_{kl}}\right\rangle^{-1} \right)
 $$
and  $v_j(d)$ for $j=1,\dots,m$ by
 $$
v_j(d)=S_{jj}-\frac{\#N_j}{d}-\sum_{k=1}^m\left\langle
S_{jk}+\frac{1}{d}\sum_{l=m+1}^n
T_{jl}\overline{T_{kl}}\right\rangle
+\frac{1}{d}\sum_{l=m+1}^n(1+\langle T_{jl}\rangle)\langle
T_{jl}\rangle\,.
 $$

\smallskip

Having constructed the approximating graph we may now investigate how the corresponding Hamiltonian behaves in the limit $d\to 0$. We denote the Hamiltonian of the star graph $G$ with the coupling (\ref{bc:st}) at the vertex as $H^\mathrm{star}$ and $H^\mathrm{approx}_d$ will stand for the approximating operators constructed above; the symbols
$R^\mathrm{star}(z)$ and $R^\mathrm{approx}_d(z)$, respectively, will denote the corresponding resolvents. Needless to say, the
operators act on different spaces: $R^\mathrm{star}(z)$ on
$L^2(G)$, while $R^\mathrm{approx}_d(k^2)$ acts on $L^2(G_d)$, where $G_d$ is the Cartesian sum $G\oplus (0,d)^{\sum_{j=1}^n N_j}$. To compare the resolvents, we identify $R^\mathrm{Ad}(z)$ with the orthogonal sum
 $$
 R^\mathrm{star}_d(z)=R^\mathrm{star}(z)\oplus0\,,
 $$
which acts as zero on the added edges. Comparing the resolvents is in principle a straightforward task, however, computationally rather demanding. Performing it we arrive at the following conclusion [\refcite{cet10a}] which provides us with the full answer to our problem on the graph level:

 \begin{theorem}
In the described setting, the operator family $H^\mathrm{approx}_d$ converges to $H^\mathrm{star}$ in the norm-resolvent sense as $d\to 0$.
 \end{theorem}

 \begin{remark}
There are various modifications of the approximation described above. In Ref.~\refcite{ct10}, for instance, the $\delta$-interactions on the connecting segments have been replaced by varying lengths of those segments; the construction is there performed for scale-invariant vertex couplings, i.e. the conditions \eqref{bc:st} with $S=0$ and any $T$.
 \end{remark}

\section{Concluding remarks}

We have demonstrated how one can use scaled Schr\"odinger operators to approximate quantum graph Hamiltonians with different vertex couplings. We have worked out the argument for the $\delta$ and $\delta'_s$-couplings. On the graph level we have provided a full solution of the problem.

This suggests how one could proceed further. The approximating graph of the previous section has to be replaced by a network with a fat edge width $\varepsilon$ and the $\delta$-couplings
by constant potentials of the appropriated strength at the segment of fat edge of length $\varepsilon$. Similarly the Laplacian is to be replaced by magnetic Laplacian on the added edges, the halflength of which is set to be $d=\varepsilon^\alpha$.  We call the resulting magnetic Schr\"odinger operator $H_\varepsilon^\omega$, where $\omega$ stands now for the
appropriate family of parameters, and by $H^\omega$ the corresponding limiting operator on the graph itself.

 \begin{conjecture}
 If $\alpha>0$ is sufficiently small the approximation result analogous to Theorem~\ref{t:delta'} is valid in the described setting for any vertex coupling \eqref{bc:st} with the same identification operator $J$.
 \end{conjecture}

Scaled potentials are not the only way how approximations of nontrivial vertex couplings can be constructed. There are other possibilities such as replacement of Neumann by suitable position dependent boundary conditions -- for a survey of fresh results we recommend Ref.~\refcite{po11}. A more difficult question is whether one can accomplish the goal by geometric means. A naive inclusion of curvature-induced potentials does not give the answer [\refcite{ep09}] a more elaborate approach has to be sought.

Let us finally comment on possible physical application of the results surveyed here. Thinking of the network as of a model of a semiconductor system, one can certainly vary the material parameters. Doping the network locally changes the Fermi
energy at the spot creating effectively a potential well or barrier. From the practical point of view, however, this does not help much because our approximations need potentials which get stronger with the diminishing tube width $\varepsilon$.

A more promising alternative is to use external fields.  In experiment with nanosystems one often adds ``gates'', or
local electrodes, to which a voltage can be applied. In this way one can produce local potentials fitting into our approximation scheme, without material restrictions. This opens an rather intriguing possibility of creating quantum graphs with the vertex coupling controllable by an experimentalist.

\section*{Acknowledgments}

The author enjoyed the pleasure of collaboration with Taksu Cheon, Olaf Post and Ond\v{r}ej Turek which led to the results reviewed here. The research was supported in part by the Czech Ministry of Education, Youth and Sports within the project LC06002.

\bibliographystyle{ws-procs9x6}

\end{document}